\documentclass[aip,jap,numerical,reprint]{revtex4-1}
\usepackage[utf8x]{inputenc}
\usepackage{amsmath,amsthm,amssymb,amsfonts}
\usepackage[usenames,dvipsnames,svgnames,table]{xcolor}
\usepackage{graphicx}
\usepackage{epsfig}
\usepackage{float}
\usepackage{afterpage}
\usepackage{verbatim}
\usepackage{preambulo}

\begin{document}
\preprint{AIP/123-QED}
\title{Phase diagram description of the CaCu$_3$Fe$_4$O$_{12}$ double perovskite}

\author{Ivon R. Buitrago}
\affiliation{Centro At\'omico Bariloche-CNEA and CONICET, Av. Bustillo Km. 9.5,  8400-Bariloche, Argentina}
\affiliation{Instituto Balseiro, Universidad Nacional de Cuyo and CNEA, 8400-Bariloche, Argentina}
\author{C. I. Ventura}
\affiliation{Centro At\'omico Bariloche-CNEA and CONICET, Av. Bustillo Km. 9.5, 8400-Bariloche, Argentina}
\affiliation{Universidad Nacional de R\'io Negro, 8400-Bariloche, Argentina}
\author{R. Allub}
\affiliation{Centro At\'omico Bariloche-CNEA and CONICET, Av. Bustillo Km. 9.5, 8400-Bariloche, Argentina}

\date{\today}

\begin{abstract}
CaCu$_3$Fe$_4$O$_{12}$ exhibits a temperature-induced transition from a ferrimagnetic-insulating phase, in which Fe
appears charge disproportionated, as \Fe{3} and \Fe{5}, to a paramagnetic-metallic phase at temperatures above
210 K, with \Fe{4} present. To describe it, we propose a microscopic effective model with two interpenetrating
sublattices of Fe$^{(4-\delta)+}$ and Fe$^{(4+\delta)+}$, respectively, being $\delta$ a measure of the Fe-charge disproportionation. We
include all $3d$-Fe orbitals: $t_{2g}$ localized orbitals, with spin 3/2 and magnetically coupled, plus two degenerate itinerant $e_g$ orbitals with local and nearest-neighbor (NN) electron correlations, and hopping between NN $e_g$ orbitals of the same symmetry. Allub and Alascio previously proposed a model to describe the phase transition
in LaCu$_3$Fe$_4$O$_{12}$ from a paramagnetic-metal to an antiferromagnetic-insulator, induced by temperature or
pressure, involving charge transfer between Fe and Cu ions, in contrast to Fe-charge disproportionation. With the model proposed for CaCu$_3$Fe$_4$O$_{12}$, modified to account for this difference between the two compounds,
the density of states of the itinerant Fe orbitals was obtained, using Green’s functions methods. The phase diagram for CaCu$_3$Fe$_4$O$_{12}$ was calculated, including phases exhibiting Fe-charge disproportionation, where the two eg orbitals in each site are symmetrically occupied, as well as novel phases exhibiting local orbital selectivity/asymmetric occupation of $e_g$ orbitals. Both kinds of phases may exhibit paramagnetism and ferromagnetism. We determined the model parameters which best describe the phase transition observed in CaCu$_3$Fe$_4$O$_{12}$, and found other phases at different parameter ranges, which might be relevant for other compounds of the ACu$_3$Fe$_4$O$_{12}$ family, presenting Fe-charge disproportionation and intersite charge transfer Fe-Cu.

\end{abstract}

\maketitle

\section{Introduction}
Double perovskites of the ACu$ _3$Fe$_4$O$_{12}$ (A: rare earth) family possess a variety of electronic and magnetic properties of interest for technological applications, e.g. as a high-activity compound for the oxygen evolution reaction that occurs during the oxidation of water, which is very important in the energy conversion reaction for metal-air rechargeable batteries~\cite{Yagi2015}, or to achieve precise control of thermal expansion coefficients~\cite{LongLa2009,LiLa2012,YamadaSr2014}

In 2007, Xiang~\emph{et al.}~\cite{XiangCa2007} studied the electronic and magnetic properties of the CaCu$_3$Fe$_4$O$_{12}$ (CCFO) double perovskite by the use of density functional theory (DFT), and they predicted that this oxide is a ferrimagnetic and half-metallic compound. In 2008 this perovskite was prepared by Yamada~\emph{et al.}~\cite{YamadaCa2008} by means of high-pressure synthesis. A phase transition at 210 K was observed from a disproportionated ferrimagnetic (FiM) state at low temperatures to a non-disproportionated or homogeneous (H) state above the critical temperature. The experimental results show a CaFeO$_3$-type charge disproportionation ($2\Fe{4}\rightarrow \Fe{3} + \Fe{5}$) in this oxide~\cite{YamadaCa2008,Shimakawa2008,MizumakiCa2011,Shimakawa2014,Shimakawa2015}.  Mizukami~\emph{et al.}~\cite{MizumakiCa2011} have reported ferromagnetic (FM) interactions between \Fe{3} and \Fe{5} spins, and antiferromagnetic (AF) coupling between Cu$^{2+}$ and Fe in this system. Recently, Yamada~\emph{et al.}~\cite{YamadaCa2016} have reported an unusual charge transfer from Fe to Cu (\emph{inverse charge transfer}) in this material. Hao~\emph{et al.}~\cite{HaoCa2009} studied the electronic and magnetic properties in this compound on the basis of density functional analysis. At the high-temperature phase they obtained a homogeneous valence (Fe$^{4+}$) and an orbital degenerate half-metallic behavior. Instead, orbital ordering, charge ordering, or disproportionation on Fe sites occur in the low temperature phase, leading to the insulating character. 
 
Since at the transition the physical properties of the material change substantially, it would be desirable to be able to control temperature and pressure. To do this it is necessary to have an understanding of the mechanism leading to the phase transition. The aim of this paper is to give a simple description of the transition, suggest a temperature-hopping (pressure) phase diagram for this compound, and describe the properties of the different phases. To this end we consider that the relevant states near the Fermi energy are mainly derived from the Fe $3d$-orbitals. Of course, these orbitals are hibridized with the Oxygen located between them, as indicated by band structure calculations. However, as a first attempt to describe this system, we eliminate the O degrees of freedom and instead consider an effective microscopic Hamiltonian taking into account the relevant effective Fe orbitals (strongly hibridized with O orbitals, as DFT calculations indicate, which would also produce a reduction of the local intra-orbital Coulomb repulsion). We include an effective Fe-Fe hopping as well as a repulsive effective nearest-neighbor density interaction. Also due to strong Hund's coupling on the Fe sites, only parallel spins would occupy the effective $\orbt$ and $\orbe$ orbitals. In order to give a simple theoretical picture we consider the orbitals of the three $\orbt$ electrons as frozen, represented by a localized spin $S_i$ and we reduce the effective $\orbe$ states to two itinerant half metallic orbitals ($x^2-y^2$ and $3z^2-r^2$) on each Fe site $i$. In this scenario, we consider four effective model parameters to characterize the physics of the system: {\bf i)} the Fe-Fe hopping integral, $t$, which favors a ferromagnetic background of the localized $S_i$ spins, {\bf ii)} the effective on-site Coulomb interaction, $\Uap$, between electrons at different $\orbe$ orbitals, {\bf iii)} the repulsive effective density interaction, $\G$, between electrons at $\orbe$ orbitals located in nearest-neighbor sites, and {\bf iv)} the magnetic coupling, $k$, between the nearest-neighbor localized $\orbt$ spins. 

As a first approach to the problem, apart from the strong interactions assumed in the infinite magnitude limit, the remaining intermediate correlations are treated using the Hartree-Fock aproximation. For one itinerant ($\orbe$) electron per site, depending on the values of the effective interaction parameters and temperature, the model describes four different phases: at high temperatures, a homogeneous paramagnetic phase (H-PM) is present and at low temperatures, with increasing $t$, the model first shows a disproportionated paramagnetic phase (D-PM), a disproportionated ferromagnetic Fe-Fe phase (D-FiM) for intermediate values of $t$, and finally a homogeneous ferromagnetic Fe-Fe phase (H-FiM) for large values of the hopping energy. We find that for some parameter ranges the proposed model provides a very reasonable description for the phase transition observed in \CCFO.

Section~\ref{sec.model} describes the model proposed for CaCu$_3$Fe$_4$O$_{12}$. Section~ \ref{sec.calculation} details the approximations and the method used to determine the phase diagram. Section~\ref{sec.results} presents our results and discusses them in the context of the relatively few previous results on this compound. Other results of the model, 
 in other parameter ranges, are discussed in Appendix A. In section~\ref{sec.conclusions} our main conclusions are presented, as well as perspectives for future investigations. 

\section{Microscopic model proposed for C\lowercase{a}C\lowercase{u}$_3$F\lowercase{e}$_4$O$_{12}$}
\label{sec.model}
In order to construct the Hamiltonian from which  to obtain the phase diagram in \CCFO, and describe the phase transition observed experimentally~\cite{YamadaCa2008} between a disproportionated ferrimagnetic insulating (D-FiM-I) phase at low temperatures and a homogeneous metallic paramagnetic (H-PM-M) phase above the critical temperature, we will propose a simplified model in which we will consider only the Fe sites and effective interactions and hopping between them.  

As a first step, to simulate the symmetric charge disproportionation (CD) between Fe ions and the NaCl-type charge ordering of the unequal Fe sites, 
experimentally observed in CCFO when the temperature is reduced~\cite{YamadaCa2008},
 we will consider that the structure of the Fe ions in CCFO is defined by two simple-cubic interpenetrated Fe sublattices that we define by Fe$^{\alpha}\equiv$\Fe{(4-\delta)} and Fe$^{\beta}\equiv$\Fe{(4+\delta)}, being $\delta$ a measure of the CD in Fe. In this way, when the CD is maximum ($\delta=1$) one has that \Fe{3} and \Fe{5} are present in the same proportion, while when the CD is minimal ($\delta=0$) only \Fe{4} is present. 

\begin{figure}[H]
\centering
\includegraphics[width=0.9\columnwidth]{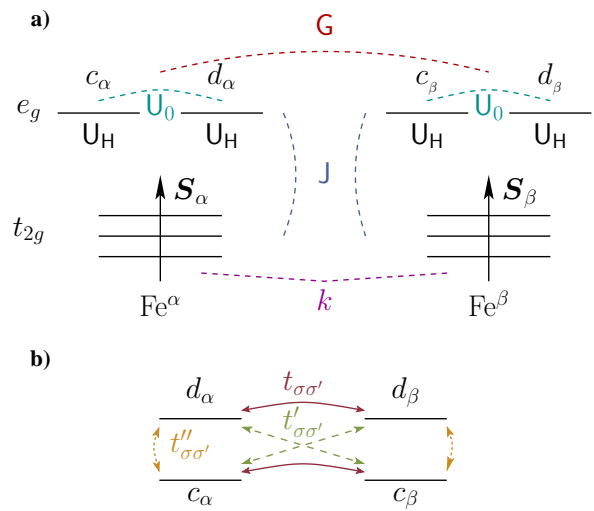}
\caption{\label{fig.model} (Color online) a) Schematic representation of interactions between localized and itinerant electrons in \Fe{\alpha} and \Fe{\beta} sublattices, and b) hopping terms between $\orbe$ electrons considered in our model.}
\end{figure}

Regarding the Fe sites, as schematized in  Figure~\ref{fig.model}, our model includes the localized $\orbt$ orbitals, represented by spins $\Sn$ with magnetic couplings ($k$) between them which might be ferromagnetic ($k=k_d$) or antiferromagnetic ($k=k_h$) as will be discussed in the next section, and two itinerant $\orbe$ orbitals per site, denoted by $c$ and $d$, with local intra- and inter-orbital electronic correlations between nearest neighbors (NN) (see Figure~\ref{fig.model}a) and three different hopping terms (see Figure~\ref{fig.model}b), depending on the symmetry of the orbitals involved. Also, we consider the Hund interaction between the spins of localized magnetic moments ($\Sn$) in $\orbt$ and those of itinerant electrons ($\sn$,$|\sn|=1/2$) in $\orbe$ orbitals, which tends to align them. 

Concretely, for itinerant electrons in the $\orbe$ bands we will consider three interaction terms: the intra-orbital Hubbard repulsion $\Ua$ between electrons of different spin in one orbital, the inter-orbital repulsion $\Uap$ between electrons $c$ and $d$ in different orbitals at one site, and the inter-orbital repulsion $\G$ between all electrons in itinerant  orbitals $\orbe$ of NN Fe-ions. Notice that for $\delta=0$, when the system is homogeneous (all Fe with total equal charge), due to the repulsion $\Uap$, the compound may present local orbital disproportionation, or orbital selectivity, corresponding to an asymmetry in the occupation of the itinerant orbitals $c$ and $d$ of an Fe site. The three hopping terms for $\orbe$ electrons are: $t_{\sigma\sigma'}$ between the same orbital $\orbe$ in neighboring sites, $t'_{\sigma\sigma'}$ between different $\orbe$ orbitals from neighboring sites, and $t''_{\sigma\sigma'}$ between $\orbe$ orbitals on the same site. 

\section{Effective Hamiltonian for C\lowercase{a}C\lowercase{u}$_3$F\lowercase{e}$_4$O$_{12}$ and phase diagram calculation.}
\label{sec.calculation}

\subsection{Effective microscopic model for CaCu$_3$Fe$_4$O$_{12}$}

To obtain the effective Hamiltonian we introduce the following approximations: {\bf i)} local intra-orbital $\Ua$ correlation much larger than $\Uap$ and $\G$. Also, infinite Hund coupling $\Jh\to\infty$, which means that the itinerant spins $\sigma$ are parallel to the localized moments $\mu$, {\bf ii)} mean-field approximation for the correlation terms $\Uap$ and $\G$, and {\bf iii)} no spin flips during hopping processes, and we neglect hopping between orbitals with different symmetry ($t'_{\sigma\sigma'}=0=t''_{\sigma\sigma'}$).

With these assumptions, the effective Hamiltonian is:
\begin{widetext}
\begin{align}
\begin{aligned}
\Ham\approx\Ham_{\sf loc} + \Ham_{\sf itin} 
- {\rm N}_s\Bigg[\frac{\widetilde{k}\,m_{\alpha}m_{\beta}}{2} + \Ge\,\nmed{\ona}\nmed{\onb}
+ \Uap\,\Big[\nmed{\nca}\nmed{\nda} + \nmed{\ncb}\nmed{\ndb}\Big] 
\Bigg]
\label{eq-Hefec}
\end{aligned}
\end{align}
\end{widetext}
where N$_s$ is the number of sites in each Fe-sublattice, $\widetilde{k}\equiv k\,z$\, is the renormalized magnetic coupling (being $z$ the coordination number), 
 $\Ge\equiv\G z$, $m_{\tau}\equiv\nmed{\Sn_{\tau}^{z}}$ is the expectation value of the z-projection of the localized moment in each sublattice $\tau$ ($\tau=\alpha,\beta$) and $\nmed{\boldsymbol{n}_{\tau\gamma}}$ is the mean occupation of the $\gamma$ itinerant orbital ($\gamma=c,d$) of sublattice $\tau$. Explicitly, Hamiltonian terms $\Ham_{\sf loc}$ and $\Ham_{\sf itin}$ are given by:
\begin{widetext}
\begin{gather}
\Ham_{\sf loc}=\sum_{i\in\tau}\bigg[\sum_{\tau'\ne\tau}
\frac{\widetilde{k}\,m_{\tau'}}{2}\,\Sn_{\tau i}^z - \Jh\big(\Sn_{\tau i}^{z}-m_{\tau}\big)\smed{\tau i}{}{z}\bigg]
\label{eq-Hloc}
\\[2mm]
\Ham_{\sf itin}=\sum_{i\in\tau}
\Big[\eatf{\tau c}\,\ncti + \eatf{\tau d}\,\ndti\Big]
-\frac{1}{2}\sum_{\nn{i}{j},\sigma} t\,\delta_{\mu_i,\mu_j}\,\Big[\casid\,\cbsj +\dasid\,\dbsj + h.c\Big]
\label{eq-Hitin}
\end{gather}
\end{widetext}
In Eq.~\eqref{eq-Hloc} $\sn_{\tau i}^z$ represents the itinerant spin operator, defined by:
\begin{gather}
\sn_{\tau i}^z=\sum_{\gamma}\frac{\on_{\tau \gamma \uparrow i}-\on_{\tau \gamma \downarrow i}}{2},
\end{gather}
while in Eq.~\eqref{eq-Hitin} $\mu$ represents the localized moment at each site $i,j$ ($\mu=+,-$), $\sigma$ represents the itinerant electron spin ($\sigma= \uparrow, \downarrow$), and the diagonal energies per site are given by:
\begin{subequations}
\begin{align}
\eatf{\tau c}&=\eat{\tau} + \Uap\nmed{\ndt} + \Ge\nmed{\ont}  \\[1mm]
\eatf{\tau d}&=\eat{\tau} + \Uap\nmed{\nct} + \Ge\nmed{\ont}
\label{eq-ener-diag}
\end{align}
\end{subequations}
being $\nmed{\boldsymbol{n}_{\tau}}$ the mean total occupation of itinerant electrons per site in sublattice $\tau$, and $\eat{\tau}$ represent the site energy of each sublattice $\tau$.

\subsection{Calculation of the phase diagram}
  
To determine the free energy from the Hamiltonian of Eq.~\eqref{eq-Hefec}, we started by the itinerant Hamiltonian contribution $\Ham_{\sf itin}$ for which we calculated the Green's functions for the electrons in itinerant orbitals $\orbe$ using the renormalized perturbation expansion RPE~\cite{Economou}, obtaining the corresponding spectral densities, $\xrho$.

The relevant phases studied are denoted: charge disproportionated (D), homogeneous in charge but with local orbital disproportionation (L) and homogeneous (H). We introduce the two following order parameters: \mbox{$\delta=0.5\big(\nmed{\ona}-\nmed{\onb}\big)$}, and $\nu=\nmed{\boldsymbol{n}_{\alpha c}}-\nmed{\boldsymbol{n}_{\alpha d}}=\nmed{\boldsymbol{n}_{\beta d}}-\nmed{\boldsymbol{n}_{\beta c}}$, 
which respectively characterize the Fe-charge disproportionation, and the local orbital disproportionation (or orbital selectivity).  
Thus, phase D corresponds to $\delta\ne 0$, $\nu=0$, while phase L is characterized by $\delta=0$, $\nu\neq 0$.
Finally, for phase H: $\delta, \nu=0$ because $\nmed{\boldsymbol{n}_{\tau \gamma}}=0.5$  for all $\tau$ and $\gamma$.

Studying the charge disproportionated and the homogeneous phases in terms of an atomic model with two NN Fe ions,~\cite{IBTesis2018}
we found that the ground states corresponding to each of these phases are, respectively, ferromagnetic (D) and  antiferromagnetic (H). Therefore, in the following we will consider the magnetic coupling constant as: $\widetilde{k}=- k_d$ for the charge disproportionated phases, while $\widetilde{k}=k_h$ for the homogeneous phases, and we exhibit the effect of each of them on the phase diagram. 

Regarding the entropy of localized spins, we consider a numerical approximation to the Brillouin entropy for \mbox{$S=3/2$} as a function of magnetization~\cite{IBTesis2018D}, similar to the one used by Allub and Alascio~\cite{AllubLa2012}. Thus, the contribution of this term  to the free energy, due to each sublattice, is then \mbox{$-T\,S_\text{loc}=-T\,\ln(4)(1-m²)^{2/3}$}, where $m$ is the normalized magnetización due to localized Fe-spins. Concretely, the free energy per site for each phase (D, L, and H)  takes the following form, respectively: 
 \begin{widetext}
 \begin{subequations}
\begin{align}
\begin{aligned}
\Fm_{\sf D}=&-k_d\,m^2-2T\ln4\big(1-m^2\big)^{2/3}
-\frac{1}{\beta} \int_{-\infty}^\infty \ln\Big[1+e^{-\beta(\omega-\mu)}\Big]\xrho_{\sf D}\big(\omega,m,\{\nmed{\nso}\}\big)d\omega
\\[1mm]&
-\Uap-2\nmed{\nca}\Big[1-\nmed{\nca}\Big]\Big[2\Ge-\Uap\Big] \label{eq-Ftotal-final-D}
\end{aligned}\\[2mm]
\begin{aligned}
\Fm_{\sf L}=&k_h\,m^2-2T\ln4\big(1-m^2\big)^{2/3}
-\frac{1}{\beta} \int_{-\infty}^\infty \ln\Big[1+e^{-\beta(\omega-\mu)}\Big]\xrho_{\sf L}\big(\omega,m,\nmed{\{\nso\}}\big)d\omega
\\[1mm]&
-\Ge-2\nmed{\nca}\Big[1-\nmed{\nca}\Big]\Uap \label{eq-Ftotal-final-L}
\end{aligned}\\[2mm]
\begin{aligned}
\Fm_{\sf H}=&k_h\,m^2 - 2T\ln4\big(1-m^2\big)^{2/3} 
- \frac{1}{\beta} \int_{-\infty}^\infty \ln\Big[1+e^{-\beta(\omega-\mu)}\Big]\xrho_{\sf H}\big(\omega,m\big)d\omega
-\frac{\Uap}{2}-\Ge \label{eq-Ftotal-final-H}
\end{aligned}    
\end{align}
\label{eq-Ftotales}
\end{subequations}
\end{widetext}
where $\beta\equiv 1/k_B T$, $\mu$ is the chemical  potencial, $\xrho$ represent the total spectral density and we have used the notation $\{\nmed{\nso}\}\equiv \{\nmed{\nca}, \nmed{\ncb}, \nmed{\nda}, \nmed{\ndb}\}$.
To determine $\mu$, in each phase we obtain the average value of the number of $\orbe$ electrons by means of the expression
\begin{gather}
\nmed{n_{\orbe}}=\int_{-\infty}^{\infty}\Big[1 + e^{-\beta\,(\omega-\mu)}\Big]^{-1}\:\xrho\big(\omega,m,\{\nmed{\nso}\}\big)\,d\omega
\label{eq-ocup-neg}
\end{gather} 
and we impose $\nmed{n_{\orbe}}=2$, as we explain below. 
By doing this self-consistently, we find the convergence values of $\mu$ and the mean occupations of each band, necessary to calculate the free energies in Eqs.~\ref{eq-Ftotales}.
To obtain the phase diagram of \CCFO\, as a function temperature (T) and hopping (t), for given values of the model parameters: $\Uap$, $\Ge$, $k_d$, and $k_h$, it is necessary to minimize the free energy of the system with respect to the magnetization (m) of the localized moments and the mean values of the occupations of the four itinerant bands $\orbe$: $\nmed{\nca}$, $\nmed{\nda}$, $\nmed{\ncb}$, $\nmed{\ndb}$, taking into account that the sum of these average occupations is 2, which corresponds to the maximum number of electrons allowed in the $\orbe$ orbitals, both in the disproportionated phase with \Fe{3}-\Fe{5} \big($\orbe^2$-$\orbe^0$\big) and  in the homogeneous phase with \Fe{4}-\Fe{4}  \big($\orbe^1$-$\orbe^1$\big). 
It is evident from Eqs.~\eqref{eq-Ftotales} that it is not possible to write an analytic form that allows us to explicitly minimize it, so that the problem must be solved self-consistently recursively, since the diagonal energies, and therefore the spectral densities, depend also on these magnitudes. For this reason the minimization of the total free energy was performed numerically, yielding the phase diagrams discussed in next section, for different model parameter ranges relevant for the description of \CCFO.

\section{Results}
\label{sec.results}


After a wide exploration of the model for different ranges of parameters, below we present the phase diagrams of the model as a function of temperature (T) and effective Fe-Fe hopping (t), with different $\Ge$ and $\Uap$. In section A the results obtained for the case $\Ge>\Uap$ are discussed. In section B we present the best fit obtained with our model to the relatively few experimental results  known for \CCFO. 
Furthermore, in Appendix A  we discuss results  obtained for the model  in the case  $\Ge<\Uap$  where phases exhibiting orbital selectivity appear, i.e. L phases,  homogeneous in charge  but with orbital disproportionation ($\nu\neq 0$) or selective occupation of the $\orbe$ orbitals, which might be of relevance for other compounds of the double perovskite family ACu$_3$Fe$_4$O$_{12}$ (A: rare earths).

In each section, the different phases obtained are analyzed and the effect of the magnetic couplings $k_h$ and $k_d$ on the phase diagrams is also discussed. For the results presented in section A and Appendix A, $\Uap$ is taken as the unit of energy. Meanwhile in section B the energies appear in eV.

\subsection{Case $\Ge>\Uap$}
\label{case1}
Let us start by considering a case $\Ge>\Uap$ such that the difference between $\Ge$ and $\Uap$ is ~10\%, i.e. $|\Ge-\Uap|=0.108\Uap$, and assume that no magnetic coupling between the localized moments exists (\mbox{$k_d=k_h=0$}). Let us mention that $\Ge>\Uap$ is reasonable, as 
$\Ge\equiv\G z$ and $z>1$, and in this situation the experimentally observed charge disproportionated ground state (D) is favoured.                                                                                                                                                                                                                                                                                                                                                                                                                                                                                                                                                                                                                                                                                                  

\begin{figure}[!htb]	
\centering
\includegraphics[width=0.9\columnwidth]{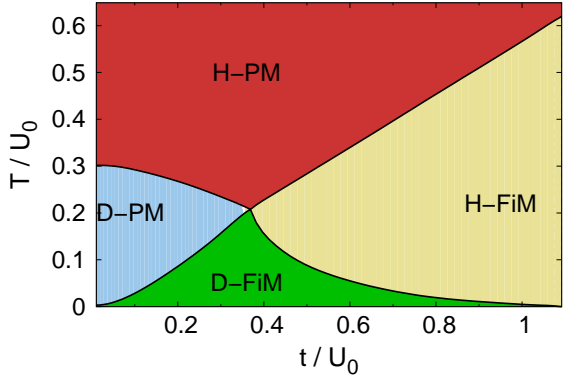}
\caption{Temperature-hopping phase diagram. Non-zero parameters: $\Uap=1$, $\Ge=1.108\,\Uap$. Phases: charge-disproportionated ferrimagnetic (D-FiM), charge-disproportionated paramagnetic (D-PM), homogeneous ferrimagnetic (H-FiM), homogeneous paramagnetic (H-PM).}
\label{fig-Diag-1} 
\end{figure}
As shown in the phase diagram of Fig.~\ref{fig-Diag-1}, four phases appear: the disproportionated ferrimagnetic (D-FiM) and the disproportionated paramagnetic (D-PM) phases that occur at low or "moderate" hopping and temperatures, and the homogeneous ferrimagnetic (H-FiM) and
homogeneous paramagnetic (H-PM) phases at larger hopping and temperature values.

For $T=0$ and very small $t$, the bands are so narrow that the kinetic energy gain of the system is negligible with respect to the repulsion $\Uap$. 


In particular, at zero hopping, because $\Ge>\Uap$, the system reduces its energy by occupying both orbitals $\orbe$ of a Fe site and leaving its neighbor empty (e.g. $\nmed{\ona}=2$ and $\nmed{\onb}=0$), and the phase is disproportionated in charge with $\delta_\text{max}=1$, as it is obtained from $\delta$ definition. As the hopping increases and the bands widen, the gain of kinetic energy begins to compensate the repulsions. In this way the charge disproportionation between Fe$^{\alpha}$ and Fe$^{\beta}$ is gradually reduced from $\delta=1$ (\Fe{3}-\Fe{5}) to $\delta=0$ (\Fe{4}-\Fe{4}), and the system passes smoothly from a disproportionated phase to a homogeneous phase, as shown in Fig~\ref{fig-Diag-1}.  
 
For $T>0$, as soon as $T$ is increased, the occupations of the bands tend to be the same and therefore ferromagnetism tends to be favored, but at the same time, due to the  magnetic entropy term,  magnetic order tends to be destroyed. This implies that, when the temperature rises, there is a competition between the FM and PM states.

\begin{figure}[!htb]
\centering
\includegraphics[width=1\columnwidth]{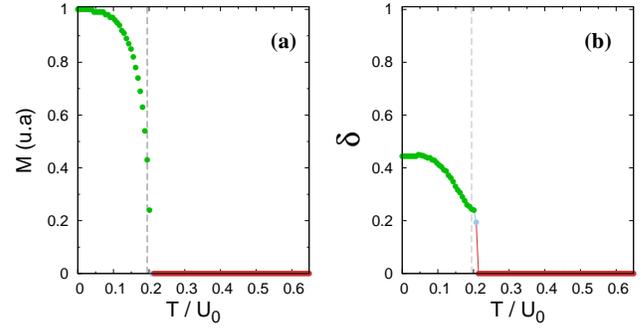}
\caption{Temperature dependence of (a) the magnetization on $\orbt$ orbitals and (b) the charge disproportionation between Fe sites when $kh=k_d=0$. Parameters: $\Ge=1.108\,\Uap$ y $t= 0.362\,\Uap$. The vertical dashed line indicates the experimental transition temperature ($T=210$ K) reported for \CCFO\, in Refs.\cite{YamadaCa2008,%
  YamadaCa2016,Shimakawa2015}.
}
\label{dMvsT}
\end{figure}

Notice in Fig.~\ref{fig-Diag-1}, that for the critical hopping value, \mbox{$t_{\sf c}\approx 0.36\,\Uap$}, the phase transition between the D-FiM and H-PM phases (experimentally observed) occurs even in the absence of magnetic coupling between the localized moments. 
At this hopping value the thermal energy, which the system acquires when the temperature rises,
it can reduce simultaneously the magnetic order of the $\orbt$ localized moments, $m$, and the charge disproportionation $\delta$ between the Fe sites, as shown in Figs~\ref{dMvsT}, which shows the temperature dependence of (a) the magnetization and (b) the charge disproportionation between Fe, respectively. 

In particular, from these two figures it can been see that: (i) the order of the localized moments in the Fe  decreases smoothly with temperature and (ii) the charge disproportionation between Fe at low temperatures it is less than its maximum value, $\delta<\delta_\text{max}=1$. These two characteristics present in our model are in agreement with what is suggested in Ref.\citep{Shimakawa2015}.

So far, the effect of the magnetic coupling between the localized moments $\orbt$ has not been considered in order to first examine under which conditions of the parameters $\Uap$ and $\Ge$ it is possible to find the type of experimental transition. However, in the most realistic case, from the experimental point of view, we should assume that these terms are non-zero and analyze their effects. As mentioned before, we will consider the magnetic coupling constant: $\widetilde{k}=-k_d$ for the charge disproportionated phases, while $\widetilde{k}=k_h$ for the homogeneous phases, and we will examine the effect of each of them  on the phase diagram of figure~\ref{fig-Diag-1} where $k_d=k_h=0$ was assumed.

\subsubsection*{Effect of $k_d$}
\begin{figure}[!htb]	
\centering
\includegraphics[width=0.9\columnwidth]{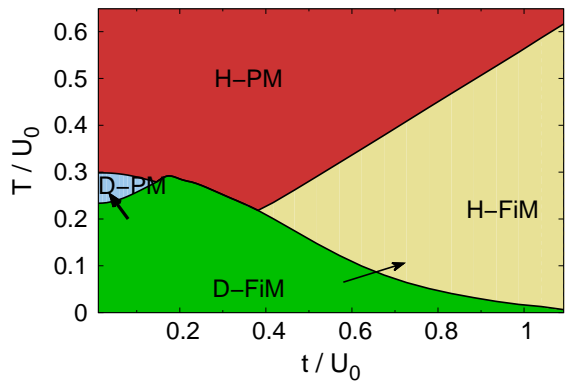}
\caption{Temperature-hopping phase diagram: effect of $k_d$. Non-zero parameters: $\Ge=1.108\,\Uap$, $k_d=0.432\,\Uap$. Arrows indicate the main effect of increasing $k_d$ on the phase diagram on the Fig.~\ref{fig-Diag-1}.}
\label{fig-Diag-3} 
\end{figure}

When considering the presence of a ferromagnetic coupling $k_d$ between the localized Fe-magnetic moments the phase diagram obtained is shown in figure~\ref{fig-Diag-3}.
Notice that: (i) the D-FiM phase is stabilized in a larger region of the phase diagram (as expected from the atomic model analysis); (ii) the effect of $k_d$ is more pronounced in the frontier between the D-FiM and D-PM phases, where the itinerant contributions to the free energies are similar, but the magnetic contribution reduces the total free energy of D-FiM phase with respect to that of the D-PM phase, and (iii) the D-PM phase is greatly reduced for \mbox{$k_d=0.432\,\Uap$} but it does not disappear. 
Although it is not shown here, from the analysis of the transition temperatures at the limit of narrow bands ($t\to0$), it can be shown that a magnetic coupling with value $k_d=\Ge-\Uap/2=0.608\,\Uap$ is necessary for this D-PM phase to disappear.
It should be noted from the phase diagram of the figure~\ref{fig-Diag-3} that taking into account only this magnetic coupling $k_d$, the hopping region where the experimental phase transition type appears is wider, and the transition temperature is reduced by increasing $t$. 
These two points would indicate that together with $\Ge$ and $\Uap$, the ferromagnetic coupling $k_d$ would be relevant when adjusting the parameters of our model to the experimental data known so far, such as the temperature of the transition $T_c$.

\subsubsection*{Effect of $k_h$}
\begin{figure}[!hbt]
\centering
\includegraphics[width=0.9\columnwidth]{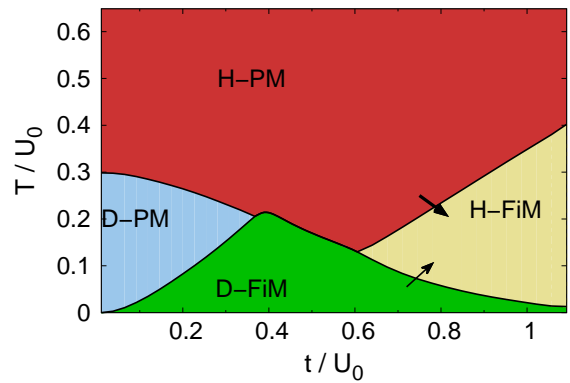}
\caption{Temperature-hopping phase diagram: effect of $k_h$. Non-zero parameters:  $\Ge=1.108\,\Uap$, $k_h=0.432\,\Uap$. Arrows indicate how the frontiers between the phases would be displaced upon increase of $k_h$.}
\label{fig-Diag-2} 
\end{figure}

When considering the antiferromagnetic coupling, $k_h$, which affects the homogeneous phases, the phase diagram obtained for $k_h=0.432\,\Uap$ is presented in figure~\ref{fig-Diag-2}. 
As seen in this diagram: (i) the stability region of the H-FiM phase is reduced with respect to the D-FiM and H-PM phases, causing the frontiers between them to move in the direction indicated by the arrows included in  Fig.~\ref{fig-Diag-2}. This is because the antiferromagnetic coupling term favors the homogeneous phases in which the $\orbt$ moments are antiparallel between them, penalizing the H-FiM phase and making it less stable; (ii) the effect of $k_h$ is less important on the frontier between the D-FiM and H-FiM phases. This occurs since in this region $\delta$ in phase D-FiM phase is appreciably reduced by the temperature so that, for each hopping value, the D-FiM phase can extend up to the temperature at which the charge distribution becomes homogeneous.



However, although it is not shown here, we confirmed that the charge disproportionation changes very smoothly near the temperature of the phase transition from D-FiM to H-PM, in contrast to the experimental facts, and therefore one would infer $k_h$ would not be so important, as we show in next section.

\subsection{Best fit to experiments in \CCFO}
We finally adjusted the model parameters so as to fit best the experimental results in \CCFO. Concretely, the critical temperature for the phase transition D-FiM to H-PM was adjusted to $\sim$ 210 K, as experimentally reported~\cite{YamadaCa2008,%
Shimakawa2014,Shimakawa2015}. At the same time, we took care to also describe the main characteristics of the temperature-dependence of the magnetization and occupation of the $3d$ orbitals reported in Ref.~\cite{Shimakawa2015}, as shown in Figs.~\ref{fig-ajustes1} and~\ref{fig-ajustes2}.

\subsubsection*{Comparison of experimental magnetization and our description.}
\begin{figure}[!htb]
\centering
\includegraphics[width=0.47\columnwidth]{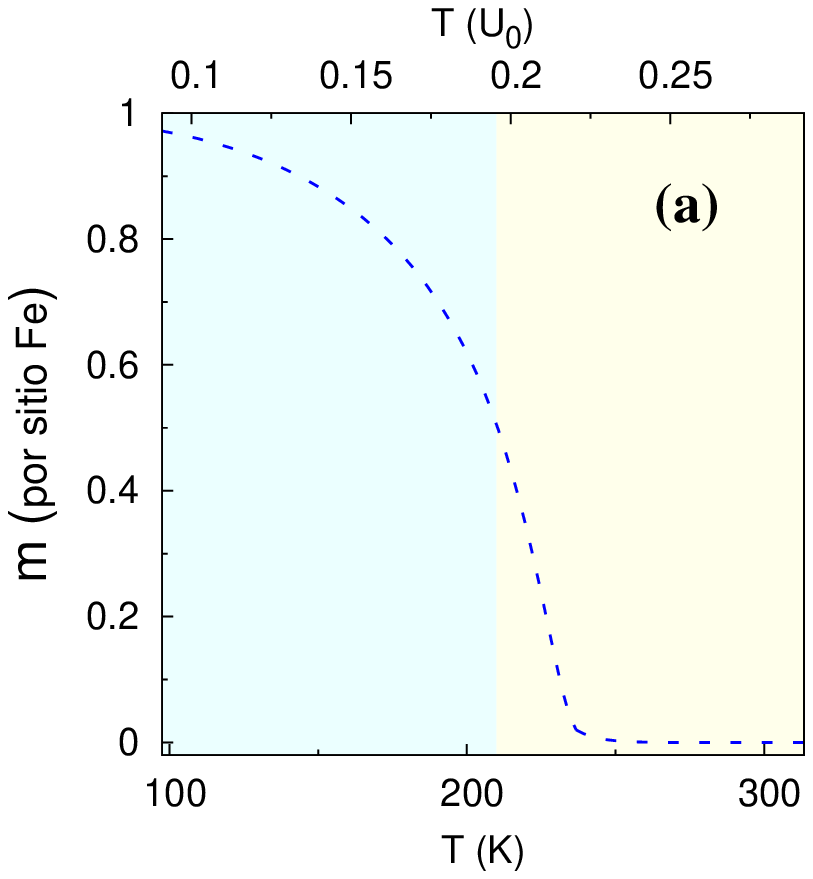}
\includegraphics[width=0.47\columnwidth]{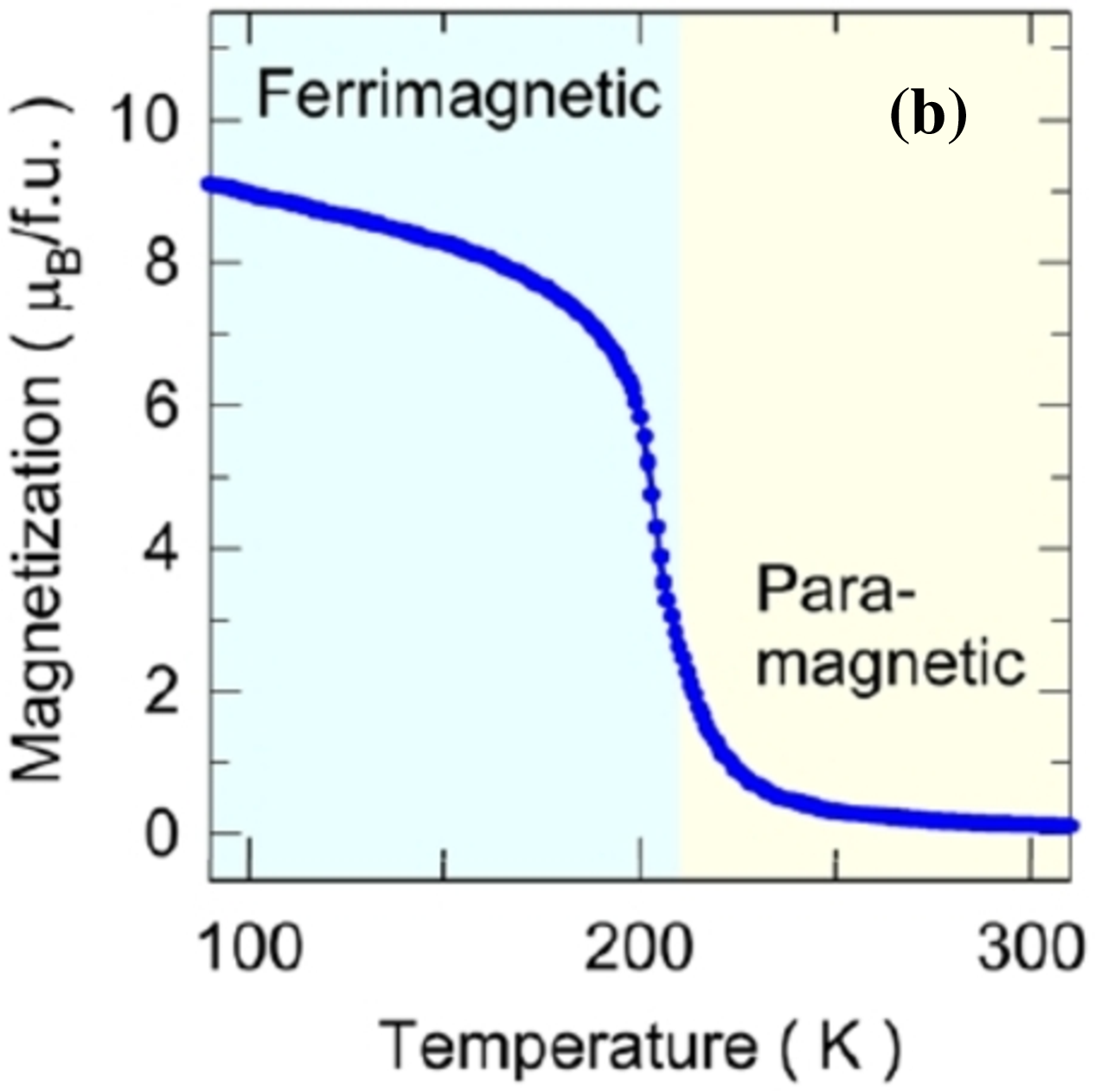}
\caption{Temperature-dependence of (a) the magnetization of moments $\orbt$ located in Fe sites obtained with our model and (b) reported by Shimakawa~\emph{et al.}~\cite{Shimakawa2015}. Parameters of our model in eV: $\Uap=0.0925$, $\Ge=0.1025$, $k_d=0.005$, $k_h=0$ and $t=0.031$. The colors in figure (a) correspond to the phases presented in the figure~\ref{fig-Diag-1} while the dotted line is  an interpolation included as a guide to the eye.}
\label{fig-ajustes1}
\end{figure}

Ref.~\cite{Shimakawa2015} presents the magnetization curve as a function of temperature, including the contributions of magnetic moments in both Fe and Cu. Although in our model we do not consider the Cu sites, in order to find the best fit, we compare the experimental magnetization  with the magnetization, $m$, per Fe site which  we obtain with our calculation. After analyzing several combinations between the four parameters of the model ($\Uap$, $\Ge$, $k_d$, $k_h$), in  figure~\ref{fig-ajustes1}a the temperature-dependence of the magnetization per Fe site, calculated with our model, is presented. The parameters considered (in eV) were \mbox{$\Uap=0.0925$}, $\Ge=0.1025$, $k_d=0.005$, $k_h=0$.
We would like to stress that the best fit is obtained for a very small value of $k_d$ and negligible $k_h$. For the hopping we consider $t=0.031$, which we found provides the optimal fit to experimental data.
Although, when comparing the Figs.~\ref{fig-ajustes1}, a qualitatively  good description of the experiment is observed, it is evident that the region where the magnetization tends to zero is narrower in our calculation (a) than in experiment (b). This could result from our calculation defining the PM phase as the one in which the magnetization of the Fe sites is strictly zero, whereas experimentally, it is seen in the Fig~\ref{fig-ajustes1}b, the condition used seems not to be so strict (within the PM phase are included points with $m\ne0$). It would seem that in Ref.~\cite{Shimakawa2015} the criterion of change in the concavity of $m$ was adopted to define the transition temperature. When taking into account that the phase defined as paramagnetic in the experiment has a magnetization whose value at 210 K ($\sim3\mub/$f.u.) corresponds to 30\% of the value of the magnetization at 90 K ($\sim 9\mub/$f.u.), these points can be assigned in our model to an H-FiM phase rather than an H-PM phase.

\subsubsection*{Temperature-dependence of total occupancies at Fe sites.}
\begin{figure}[!h]
  \centering
    \includegraphics[width=0.49\columnwidth]{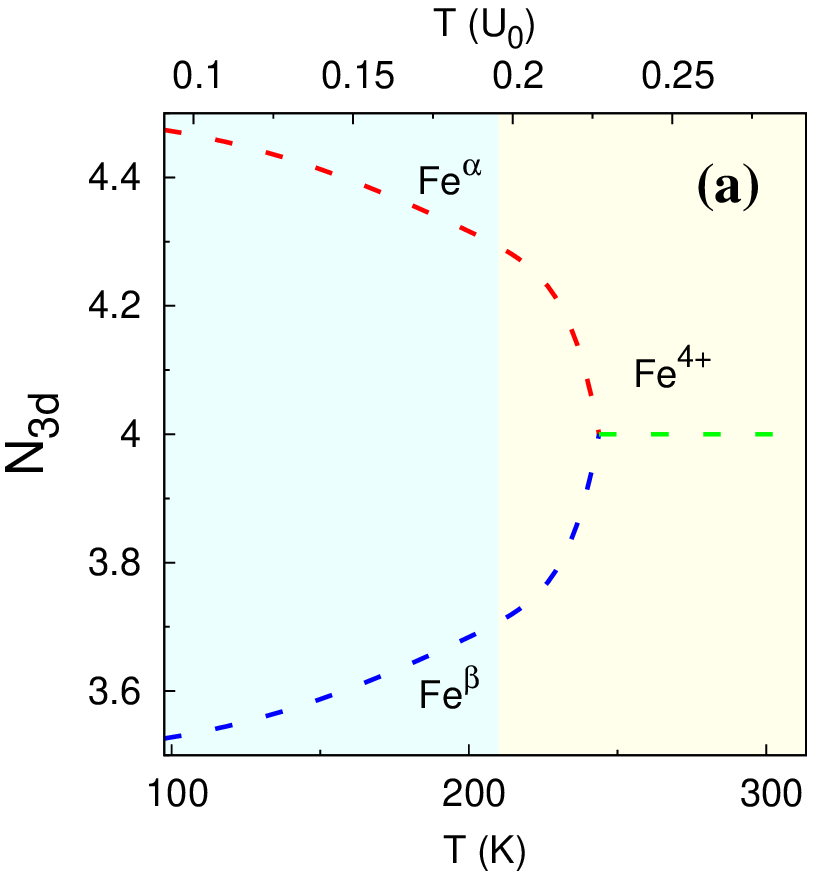}
    \includegraphics[width = 0.49\columnwidth]{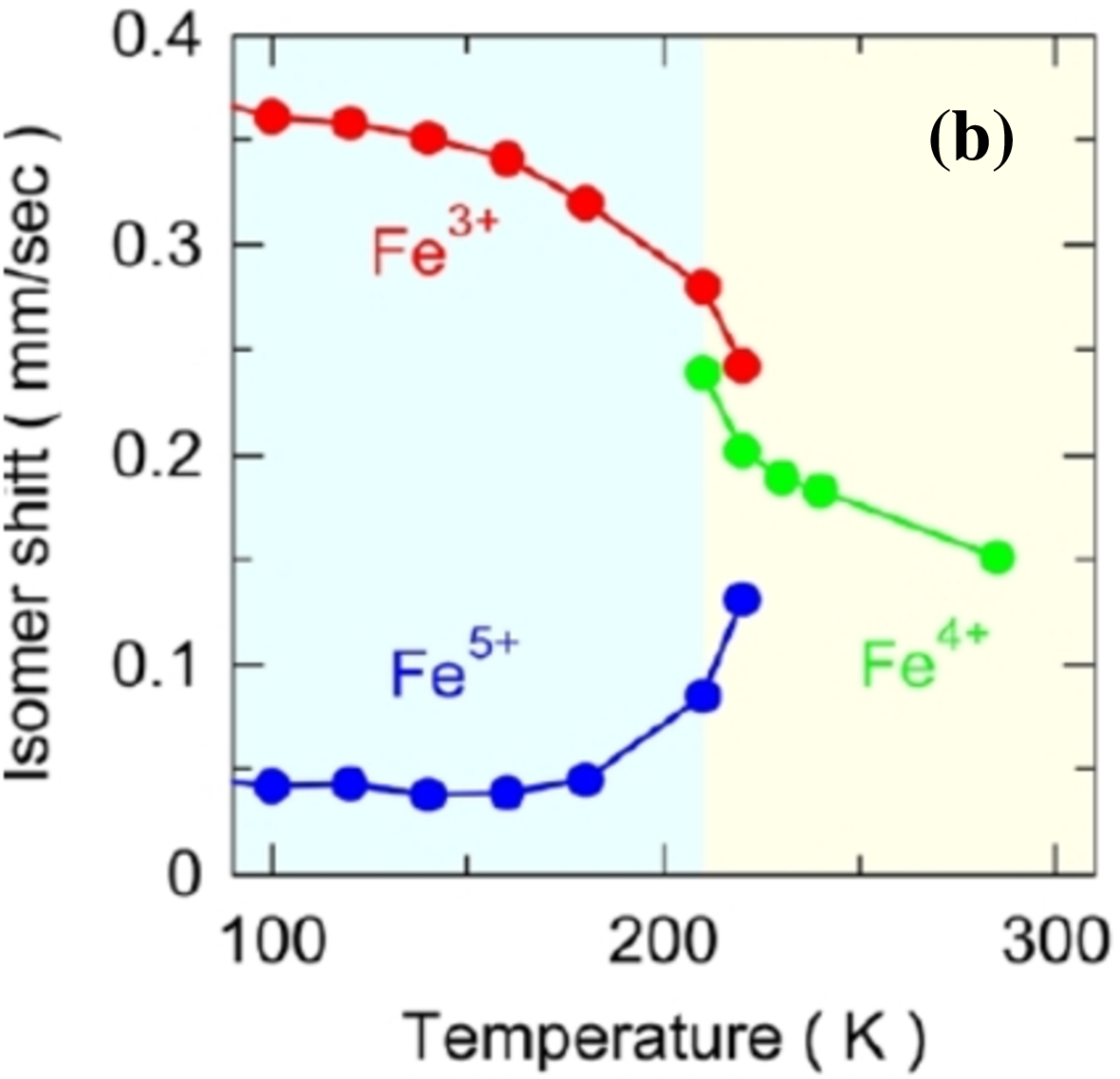}
\caption{Temperature-dependence of (a) the total occupations of the $3d$-Fe orbitals calculated in this work and (b) the isomeric shift reported in Ref.~\cite{Shimakawa2015}. The parameters are the same as in Fig.~\ref{fig-ajustes1}. The colors in (a) correspond to the phases presented in the Fig~\ref{fig-Diag-1} and the dotted line is an interpolation to calculated data included as a guide to the eye.}
  \label{fig-ajustes2}
\end{figure}

Now we will compare the temperature-dependence of the total occupancies in the \mbox{$3d$-Fe} orbitals, obtained with our model, and the isomeric shift curve reported in Ref.~\cite{Shimakawa2015}. From the assumptions of the model the total occupations of the $3d$-Fe orbitals can be calculated by means of the following expressions,
\begin{subequations}
\begin{align}
N_{3d}^{\;\alpha}=n_{\orbt}^{\;\alpha} + n_{\orbe}^{\;\alpha}
= 3 + \nmed{\nca} + \nmed{\nda} \\[2mm]
N_{3d}^{\;\beta}=n_{\orbt}^{\;\beta} + n_{\orbe}^{\;\beta}
= 3 + \nmed{\ncb} + \nmed{\ndb}
\end{align}
\label{eq-Ntotal}
\end{subequations}
where the average values ​​$\nmed{\boldsymbol{n}_{\tau\,\gamma}}$ are those that minimize the free energy for each value of hopping and temperature. In the Fig.~\ref{fig-ajustes2} is shown the temperature-dependence of the occupations of the $3d$-Fe orbitals (Eqs.~\eqref{eq-Ntotal}) corresponding to the parameters of the Fig.~\ref{fig-ajustes1}.

The difference in scale between the data in Figs.~\ref{fig-ajustes2} is due to the fact that the magnitudes plotted are not exactly the same: the isomeric shift is connected to the valence at each Fe$^{\alpha}$ and Fe$^{\beta}$ site, as indicated in figure~\ref{fig-ajustes2}b, and therefore is also related to the total occupations of $\orbe$ orbitals. Anyway, when comparing both Figs.~\ref{fig-ajustes2}, it is seen that our model qualitatively reproduces the changes with temperature of the charge disproportionation $\delta$ that experimentally shows the isomeric shift.

\begin{figure}[!h]
\includegraphics[width=1\columnwidth]{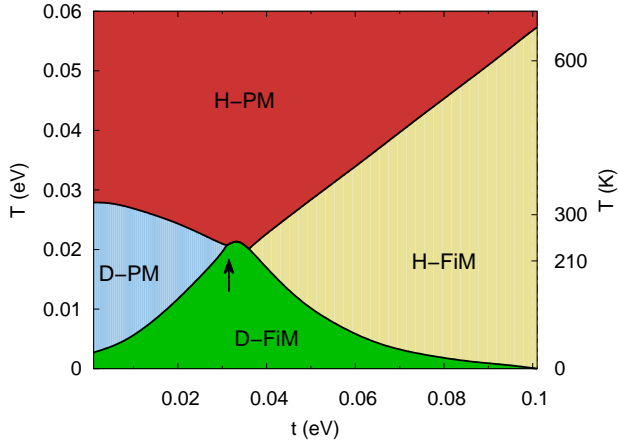}
\caption{Phase diagram $T$-$t$ of our proposed model for \CCFO, with model parameters providing the best fit to experimental data. Parameters (in eV): $\Uap=0.0925$, $\Ge=0.1025$, $k_d=0.005$, $k_h=0$.}
\label{fig-Diag-7}
\end{figure}

Notice here that, according to the experimental results of the Fig.~\ref{fig-ajustes2}b, there is a region around the transition temperature in which the three valences (\Fe{3}, \Fe{5}and \Fe{4}) seem to coexist, which would indicate that the phase transition between  D-FiM and H-PM is not so sharp. 
This fact, together with the Fig.~\ref{fig-ajustes1}b, would suggest that the phase transition  
experimentally reported as: D-FiM $\to $ H-PM for \CCFO\, and which our model describes for a range of hopping values around $t=0.031$ eV,
as indicated by the arrow in Fig.~\ref{fig-Diag-7}, might perhaps under pressure occur also passing  through different intermediate states (for intermediate temperatures), according to one of the following two sequences of phase transitions:
\begin{center}
\begin{tabular}{cccccc}
I: & D-FiM & $\to$ & D-PM & $\to$ & H-PM \\
II:& D-FiM & $\to$ & H-FiM & $\to$ & H-PM
\end{tabular}
\end{center}
Both possibilities appear in our phase diagram of  Fig.~\ref{fig-Diag-7}: sequence I (through an intermediate D-PM phase) is possible at lower hopping values (decreasing  pressure), while sequence II (through an intermediate H-FiM phase)  is possible at  larger hopping values (increasing pressure).

It would be interesting to test these predictions with experimental investigations of the phase diagram of  \CCFO\, under pressure.

\begin{figure}[!h]
  \centering
    \includegraphics[width=1\columnwidth]{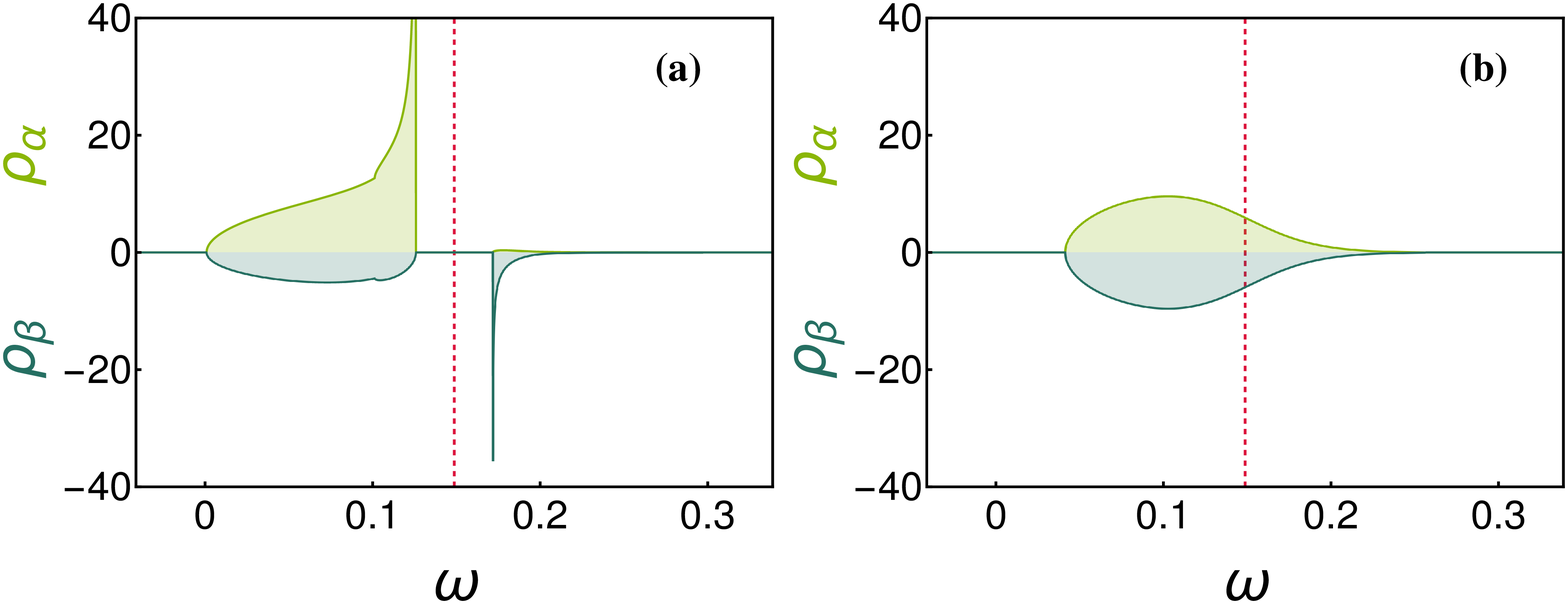}
   \caption{Total spectral densities of both $\orbe$-Fe orbitals 
   for (a) D-FIM phase at $T=160.2$ K which correspond to $m=0.85$, and (b) H-PM phase at $T=250.7$ K with $m=0$. Parameters (in eV): $\Uap=0.0925$, $\Ge=0.1025$, $k_d=0.005$, $k_h=0$ and $t=0.0310$.}
  \label{fig-densities}
\end{figure}

Regarding the metallic character of each phase, in the figure~\ref{fig-densities} we present the total spectral densities for both orbitals $\orbe$-Fe, obtained with our model in points (a) \mbox{$t=0.031$ eV} and $T=0.0138$ eV ($\sim 160.2$ K) within the D-FiM phase, and (b) $t=0.031$ eV and \mbox{$T=0.0216 $ eV} ($\sim 250.7$ K) within the H-PM phase. As it is appreciated, the energy gap in the spectral density of the D-FiM phase would allow to attribute to it a semicondutor/insulator character, while the H-PM phase would be metallic, as the experiments suggest~\citep{YamadaCa2008,Shimakawa2015}.

Let us now analyze the charge disproportionation obtained with the model parameters providing the  best fit to experiments.  
Taking into account  Fig.~\ref{fig-ajustes2}a, from our  definition of symmetric charge disproportionation at Fe sites $\delta$,
\begin{gather}
\delta\equiv |\nmed{\nca}-\nmed{\ncb}|=\frac{\Big|\nmed{\ona} - \nmed{\onb} \Big|}{2}=\frac{\big|N_{3d}^{\: \alpha} -N_{3d}^{\: \beta} \big|}{2},
\end{gather}
At  90 K  we can thus  estimate $\delta_{\rm mod} (90{\rm K})\approx 0.5$, so that one would have: 
\begin{align*}
\text{Fe}^{\alpha}&\equiv \Fe{(4- \delta)}=\Fe{3.5}   \\
\text{Fe}^{\beta} &\equiv \Fe{(4+\delta)}=\Fe{4.5}
\end{align*}

\section{Conclusions}
\label{sec.conclusions}
We propose the first microscopic effective model which describes the phase transition experimentally observed in \CCFO, between the charge disproportionated ferromagnetic phase (D-FiM) and the homogeneous paramagnetic phase (H-PM). The model includes two interpenetrated Fe sublattices, with two $\orbe$ itinerant electron orbitals per site, possessing local intra-orbital and inter-orbital correlations, as well as nearest-neighbour correlations. Also magnetic coupling is considered between the localized magnetic moments representing  $\orbt$ orbitals, as well  as Hund coupling  between the  itinerant and localized spins.

With the analytical treatment performed for the simplified model proposed, we were able to: 
\begin{itemize}
\item[i)]  obtain the phase diagram of the model, which includes the phases experimentally observed in \CCFO,   and analize the effect of the different correlations and magnetic couplings;
\item[ii)] obtain a range of model parameters which allows to describe the experimentally observed phase transition, mainly driven in this model by the 
nearest-neighbour correlations $\Ge$ between itinerant electrons;
\item[iii)] we could also describe the temperature-dependence of the magnetization and the charge disproportionation between Fe reported for \CCFO.
\item[iv)] the model also predicts some phases not yet experimentally observed, like the disproportionated paramagnetic (D-PM) and homogeneous paramagnetic (H-FiM) phases, which could be found in experiments under pressure . Interestingly, in another parameter range ($\Ge<\Uap$), also a phase (L) with selective occupation of the$\orbe$ orbitals is obtained (corresponding to$\nu \neq 0$): this orbital selectivity might be relevant for other compounds of the double perovskite family ACu$_3$Fe$_4$O$_{12}$ (A: rare earths), which are believed to present varying degrees of  charge disproportionation and/or charge transfer.  For $\Ge>\Uap$ and $\Ge<\Uap$ the  model exhibits a symmetry
 between the order parameters  $\delta$ and $\nu$,  respectively measuring charge disproportionation and orbital selectivity,  reflected in the corresponding  phase diagrams.
\end{itemize}

Having analyzed the model for cases $\Ge>\Uap$ and $\Ge<\Uap$, it is observed that  \CCFO\, is best described by assuming:  $\Ge>\Uap$, and  the origin 
of the experimentally reported  D-FiM to H-PM phase transition is connected to the nearest-neighbor electron correlation $\G$ in our approach. 
 However, the analysis of the $\Ge<\Uap$ case is still important and could be relevant to explain future experiments in other double perovskite compounds of this family. 
 
At this point, it is important to emphasize that  this simplified model and analytical  treatment  provide  suitable approach to explain the phase transition  reported in \CCFO, and could be used as a starting point to formulate a more general theory that makes it possible to explain the experimental phase diagrams recently reported by Chen~\emph{et al.}~\cite{ChenCaLa2012} for the solid solutions La$_x$Ca$_{1-x}$Cu$_3$Fe$_4$O$_{12}$ with $x=0, \xfrac{1}{4}, \xfrac{1}{2}, \xfrac{3}{4}, 1$, as well as, other members of the \ACFO\, family\cite{EtaniY2013,YamadaCe2014,%
MurakamiCaYCe2016,YamadaCaSr2017,%
YamadaLn2013}, which can present both, intersite charge transfer Cu-Fe  like for $ x = 1 $ and Fe-charge disproportionation like for $ x=0 $, here  (but with differents ratios \Fe{3}/\Fe{5} as reported for example in \SCFO\,~\cite{YamadaSr2011,YamadaSr2014,%
YamadaSr2016}), and other features that make their study  more complex, not yet theoretically done yet.

\section*{Acknowledgments}
We are especially grateful to Prof. B. Alascio for the motivating discussions and his valuable comments and advice. We acknowledge financial support by CONICET (PIP Grant 0702, and the fellowship awarded to I.R.B.). C.I.V. and R. A. are members of Carrera del Investigador Cient\'\i fico, CONICET.

\appendix
\section{Phase diagrams for  $\Ge<\Uap$}
\label{case2}
We already shown that when considering $\Uap<\Ge$ it is possible to find the phase transition from D-FiM to H-PM reported in several experiments ~\cite{YamadaCa2008,%
MizumakiCa2011, Shimakawa2015}. However, we found that it is possible to find average occupations corresponding to the disproportionate phase, provided that $\Uap\leqslant 2\Ge$, now we will consider the case in which $\Uap/2<\Ge<\Uap$ and analyze the phase transitions possible in this case for the proposed model.

\begin{figure}[!htb]
\centering
\includegraphics [width=0.9\columnwidth]{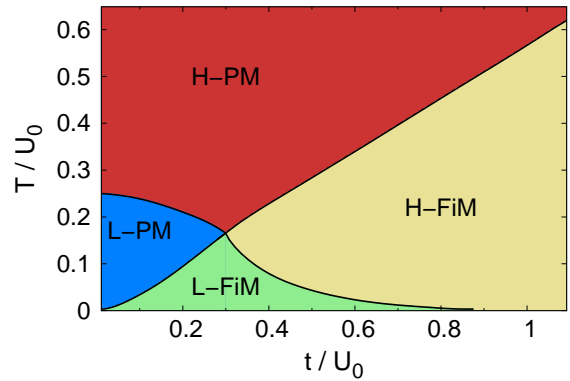}
\caption{Phase diagram $T$-$t$ for case $\Ge=0.892\,\Uap$ without magnetic couplings. Phases: homogenous in charge with local orbital disproportionation ferrimagnetic (L-FiM), homogeneous in charge with local orbital disporporcionacion paramagnetic (L-PM), homogeneous ferromagnetic (H-FiM), homogeneous paramagnetic (H-PM).}
\label{fig-Diag-4}
\end{figure}

To start, in figure~\ref{fig-Diag-4} we present the phase diagram obtained for $\Ge=0.892\,\Uap$ and $k_h=k_d=0$. We have chosen this value for $\Ge$ since it corresponds to the symmetric case of the diagram in Fig.~\ref{fig-Diag-1}, that is, both have the same difference $|\Ge-\Uap| =0.108\,\Uap$.
In this case, in addition to the H-FiM and H-PM phases, there is also the phases homogeneous in charge with orbital disproportionation ferromagnetic (L-FiM) and the homogenous in charge with orbital disproportionation paramagnetic (L-PM). Clearly, the phase diagram is qualitatively similar to that obtained with $\Ge= 1.108 \,\Uap$, in terms of the presence of H-FiM and H-PM phases in similar hopping and temperature ranges. However, the main difference with the diagram in Fig.~\ref{fig-Diag-1} is that the charge-disproportionated phases  D-FiM and D-PM  appear replaced by phases with local orbital selectivity in each Fe:  the L-FiM and L-PM phases. In this case, because $\Ge<\Uap$, for low hopping and temperature the system reduces its energy by paying $\Ge$ instead of $\Uap$. This is achieved by occupying more of one of the two bands $\orbe$ in each site ($\nu\ne0$) whereas maintaining the same charge among the Fe sites ($\delta=0$). As in the $\Ge>\Uap$ case, as the hopping increases the effect of orbital disproportionation $\nu$ is compensated by the energy that the system gains when 
forming bands, and therefore, the occupations of both $\orbe$ orbitals tend to gradually homogeneize until for  $t\sim0.9 \,\Uap$ they are equal.

The above results allow us to highlight the symmetry that exists between $\delta$ and $\nu$ with respect to the role they play in the phase diagrams when magnetic couplings are not considered. Depending on who is larger, $\Ge$ or $\Uap$, the phases at low temperatures and hopping are D-FiM \big($\Ge>\Uap$\big) or L-FiM \big($\Uap/2<\Ge<\Uap$\big).

In what follows we study the effect of magnetic couplings on the diagram~\ref{fig-Diag-4}. In particular, we will analyze under what conditions the D-FiM phase can arise and we will also examine if it is feasible to find the experimental phase transition for this relationship between $\Ge$ and $\Uap$ with reasonable values for the magnetic couplings.

\subsubsection*{Effect of $k_h$}
\begin{figure}[!hbt]
\centering
    \includegraphics[width=0.9\columnwidth]{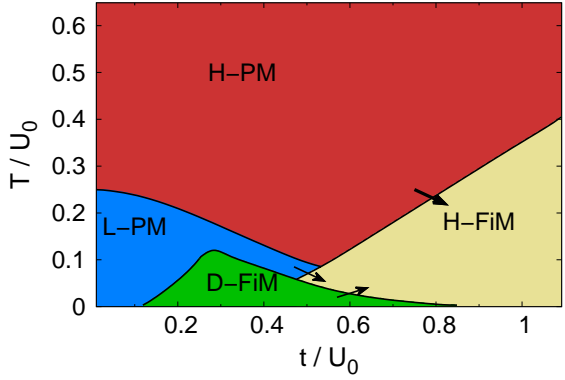}
\caption{Phase diagrams $T$-$t$ for case $\Ge= 0.892\,\Uap$ with antiferromagnetic coupling $k_h=0.430\,\Uap$. The arrows indicate the direction in which the boundaries between phases move respect to Fig.~\ref{fig-Diag-4} when $k_h$ increases.}
\label{fig-Diag-5}
\end{figure}

When considering the antiferromagnetic coupling $k_h=0.483\,\Uap$ the phase diagrams obtained is shown in  figure~\ref{fig-Diag-5}.

As seen in this phase diagram, the presence of the magnetic energy term increases the free energy of the L-FiM and H-FiM phases with respect to the others, and therefore, the D-FiM phase becomes stable in the region where before  the L-FiM phase was stable. Note that, for this relationship between $\Ge$ and $\Uap$, there is always a region of the diagram ($t\lesssim0.12\,\Uap$) where the stable phase at $T=0$ is paramagnetic (L-PM).

In addition, by increasing $k_h$ the borders between the phases move in the directions indicated by the arrows in the figure~\ref{fig-Diag-5} and the phase D-FiM is confined to a small region, being impossible to find the transition from phase D-FiM to H-PM. When comparing the transition temperatures, obtained from the free energy in the limit of narrow bands, we find that the phase transition from D-FiM to H-PM occurs whenever the ferromagnetic coupling constant $k_d\sim0.608\,\Uap$. As in the diagram~\ref{fig-Diag-5} the ferromagnetic coupling $k_d$ is zero, this explains why it is not possible to obtain a reasonable phase diagram (with magnetic phases at $T=0$), nor find the experimental phase transition for some hopping including only antiferromagnetic coupling $k_h$.

\subsubsection*{Effect of $k_d$}
\begin{figure}[!htb]
\centering
 \includegraphics[width=0.9\columnwidth]{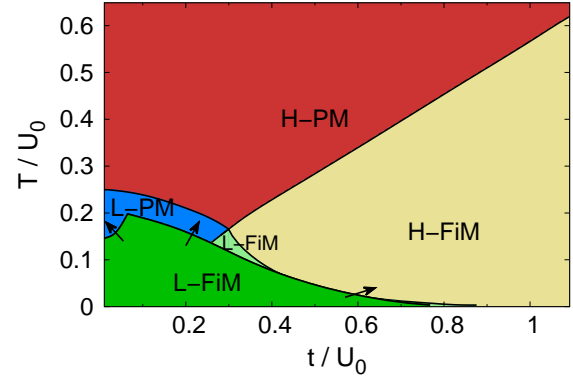}
\caption{Phase diagram $T$-$t$ for case $\Ge=0.892\,\Uap$ with ferromagnetic coupling $k_d=0.430\,\Uap$.}
\label{fig-Diag-6}
\end{figure}

As we explained in the previous paragraph, it is necessary to include the ferromagnetic coupling $k_d$ adequately so that, in the limits of narrow bands ($t\to0$) and low temperatures ($T\to0$), the stable phase will be ferromagnetic (D-FiM). When considering  $k_d=0.430\,\Uap$ the obtained phase diagram is presented in the figure~\ref{fig-Diag-6}.

As seen in Fig.\ref{fig-Diag-6} the ferromagnetic coupling $k_d$ reduces the free energy of the D-FiM phase and it begins to become stable in an increasingly wider region, where previously were the L-FiM and L-PM phases. In particular, when $k_d>0.108\,\Uap$ the phases at $T=0$ are always FM as is to be expected since the localized $\orbt$ moments are frozen. However, as the value of $k_d$ increases, as for the diagrams with $k_h\neq0$ the D-FiM phase is always confined between phases L-FiM and L-PM, at least with the values for the magnetic coupling constants assumed. This would seem to contradict what we mentioned above, that in the narrow band limit the phase transition from D-FiM to H-PM should occur when $k_d\geqslant0.608\,\Uap$. 

However, if it is taken into account that to find the transition temperatures in this limit it was assumed that $T \ll |\omega-\eat{f}|$, as the temperature rises, this condition is no longer fulfilled and therefore the value of the critical $ k_d $ could be overestimated. In addition, since a greater value of $k_d$ higher is the magnetization of the D-FiM phase and smaller the gap that must be overcome thermally to homogeneize the occupations of the bands, the transition temperature between phases D-FiM and L- PM may also be overestimated. Finally, although it is not shown here, we have verified that by including both magnetic couplings $k_h$ and $k_d$ simultaneously, the phase transition from D-FiM to H-PM is not found. 

In summary, having analyzed the cases $\Ge>\Uap$ and $\Ge<\Uap$, it is observed that for \CCFO\, it is convenient to always consider the first one. However, the analysis of the $\Ge<\Uap$ case is still important and could be relevant to explain future experiments in other family members. Furthermore, in this model the origin of the D-FiM to H-PM phase transition is connected to the nearest-neighbor electron correlation$\G$.


\bibliographystyle{unsrt}
\bibliography{biblio}

\end{document}